\documentclass[fleqn,12pt,twoside]{article}
\usepackage[headings]{espcrc1}

\readRCS
$Id: espcrc1.tex,v 1.2 2004/02/24 11:22:11 spepping Exp $
\usepackage{graphicx}
\usepackage[figuresright]{rotating}

\newcommand{\AmS}{{\protect\the\textfont2
  A\kern-.1667em\lower.5ex\hbox{M}\kern-.125emS}}

\title{Rapidity and system size dependence of anisotropic flows in
relativistic heavy ion collisions}

\author{Che Ming Ko\address{Cyclotron Institute and Physics Department,\\
    Texas A\&M University, College Station, Texas 77843-3366, USA}%
    \thanks{Work supported by US National Science Foundation under
    Grant No. PHY-0457265 and the Welch Foundation under Grant
    No. A-1358.},
       Lie-Wen Chen\address{Institute of Theoretical Physics,\\
    Shanghai Jiao Tong University, Shanghai 200240, China}%
    \thanks{Work supported by National Natural Science Foundation of China
    under Grant Nos. 10105008 and 10575071.}}

\runtitle{Rapidity and system size dependence of anisotropic flows}
\runauthor{C. M. Ko and L. W. Chen}

\begin{document}

\maketitle

\begin{abstract}

We report results from a multiphase transport (AMPT) model on the
rapidity and system size dependence of charged hadron anisotropic flows 
in nuclear collisions at the Relativistic Heavy Ion Collider (RHIC). 
\end{abstract}

\section{Introduction}

The azimuthal anisotropy of hadron transverse momentum
distributions in heavy ion collisions at RHIC is sensitive to the
properties of produced matter. In the framework of transport
approach, it was shown in Ref.\cite{Zhang99} using the ZPC 
model \cite{Zhang:1997ej} that the value of the second harmonic, 
i.e., the elliptic flow, depends sensitively on the magnitude 
of parton scattering cross sections. With a more realistic 
collision dynamics via the AMPT model \cite{Zhang:2000bd,Lin:2001cx}, 
it was further shown that the observed large elliptic flow and 
its ordering according to hadron masses \cite{Lin:2001zk} as well 
as high-order anisotropic flows \cite{chen1} could be explained if 
partons scatter with cross sections much larger than those given 
by the perturbative QCD. Including also charm quarks in the AMPT model
\cite{zhang}, the observed large elliptic flow of electrons from
the decay of charmed mesons is again consistent with a large charm
scattering cross section. In the present contribution, recent
results from the AMPT model on the rapidity and collision system
dependence of anisotropic flows are presented
\cite{chen2,chen3,chen4}.

\section{The AMPT model}

The AMPT model is a hybrid model that uses minijet partons from 
hard processes and strings from soft processes in the HIJING model 
\cite{Wang:1991ht} as the initial conditions for modelling heavy ion 
collisions at ultra-relativistic energies. In the default version, 
time evolution of resulting minijet partons is described by the ZPC 
model \cite{Zhang:1997ej} with an in-medium cross section derived 
from the lowest-order Born diagram with an effective gluon screening 
mass taken as a parameter for fixing the magnitude and angular 
distribution of parton scattering cross section. After minijet 
partons stop interacting, they are combined with their parent 
strings, as in the HIJING model with jet quenching, to fragment 
into hadrons using the Lund string fragmentation model as implemented 
in the PYTHIA program \cite{Sjostrand:1994yb}. The final-state 
hadronic scatterings are then modelled by the ART model 
\cite{Li:1995pr}. In an extended string melting version of the
AMPT model \cite{Lin:2001zk}, hadrons that would have been
produced from string fragmentation are converted to valence quarks
and/or antiquarks in order to model the initially formed partonic
matter. Interactions among these partons are again described by 
the ZPC parton cascade model. The transition from the partonic matter 
to the hadronic matter is achieved using a simple coalescence model, 
which combines two nearest quark and antiquark into mesons and 
three nearest quarks or antiquarks into baryons or anti-baryons that 
are close to the invariant mass of these partons.

\section{Pseudorapidity dependence of anisotropic flows}

\vspace{-0.5cm}

\begin{figure}[htb]
\begin{minipage}{18pc}
\includegraphics[scale=0.75]{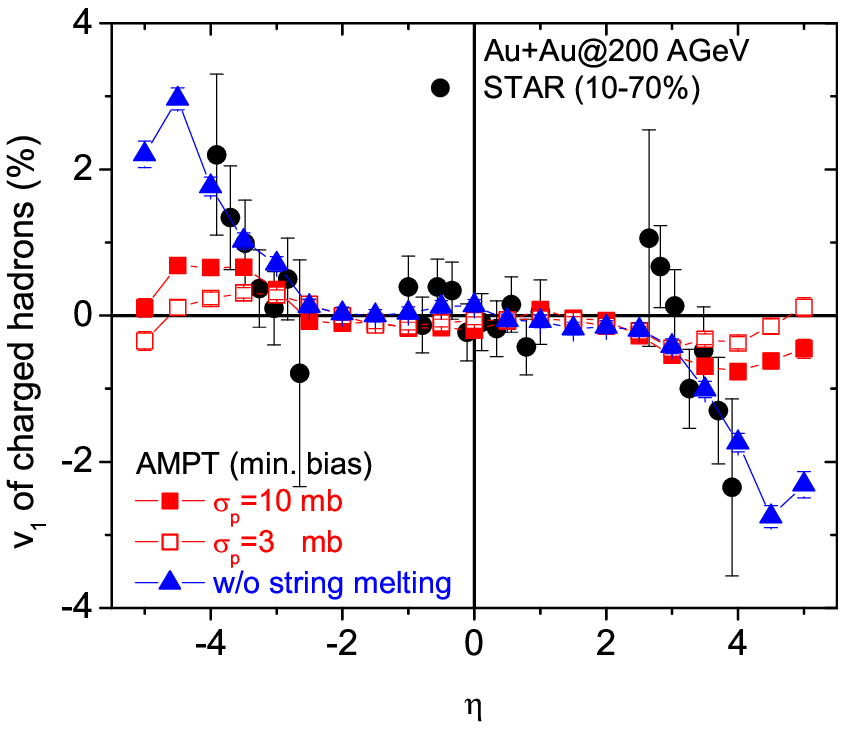}
\end{minipage}
\begin{minipage}{18pc}
\includegraphics[scale=0.75]{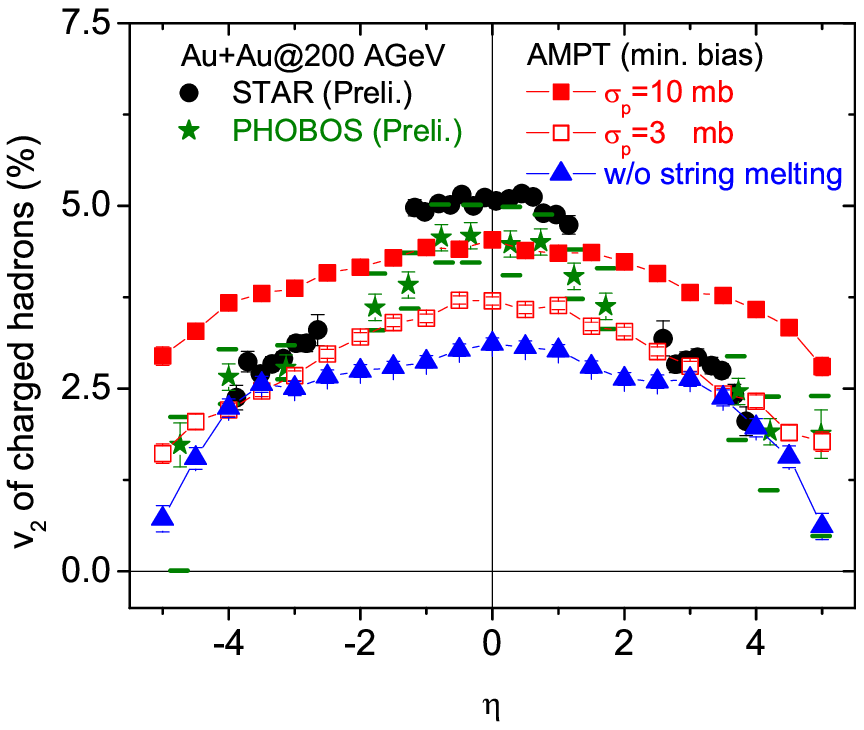}
\end{minipage}
\vspace{-0.5cm}
\caption{Pseudorapidity dependence of $v_{1}$ (left panel) and $v_2$
(right panel) in minimum bias events of Au + Au collisions at
$\sqrt{s}=200$ AGeV.}
\label{rapidity}
\end{figure}

Results from the AMPT model on the pseudorapidity ($\eta$) dependence 
of the directed ($v_{1}$) and elliptic ($v_2$) flows of charged hadrons 
in minimum bias events of Au + Au collisions at $\sqrt{s}=200$ AGeV 
are shown, respectively, in the left and right panels of 
Fig. \ref{rapidity} \cite{chen2}. For $v_1$, both the default version 
and the version with string melting can reproduce approximately the 
STAR data (solid circles) \cite{STAR03} around the mid-pseudorapidity 
region, while only the default version can describe the data at 
large $\left\vert \eta \right\vert$. For $v_2$, the string melting 
scenario with a parton scattering cross section of 
$\sigma _{p}=$ $10$ mb (solid squared) describes very 
well the PHOBOS data (solid stars) \cite{manly03} around mid-$\eta $ 
($\left\vert \eta \right\vert \leq 1.5$) but a smaller
$\sigma_{p}=3$ mb (open squares) or the default version (solid
triangles) gives a better description of both PHOBOS and STAR
\cite{oldenburg04} data at large pseudorapidity
($\left\vert \eta \right\vert \geq 3$). These interesting features 
may imply that initially the matter produced at large pseudorapidity
is dominated by strings while that produced around mid-rapidity
mainly consists of partons. This is a reasonable picture as particles
at large rapidity are produced later in time when the volume of the
system is large and the energy density is small.

\section{System size dependence of anisotropic flows}

\vspace{-0.5cm}

\begin{figure}[htb]
\begin{minipage}{18pc}
\includegraphics[scale=0.65]{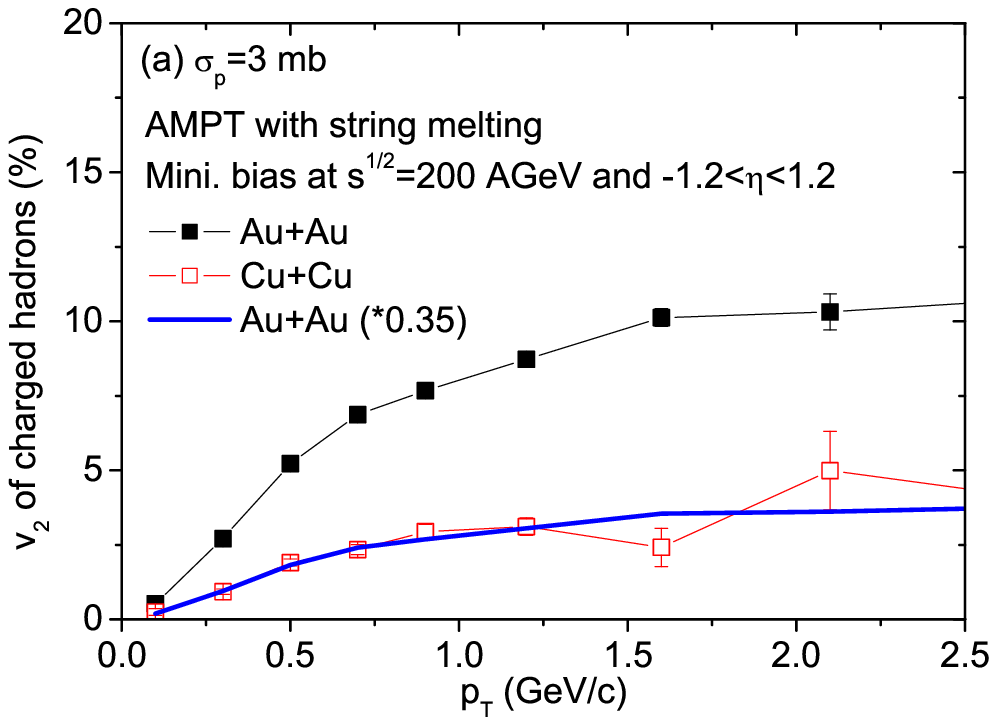}
\end{minipage}
\begin{minipage}{18pc}
\includegraphics[scale=0.65]{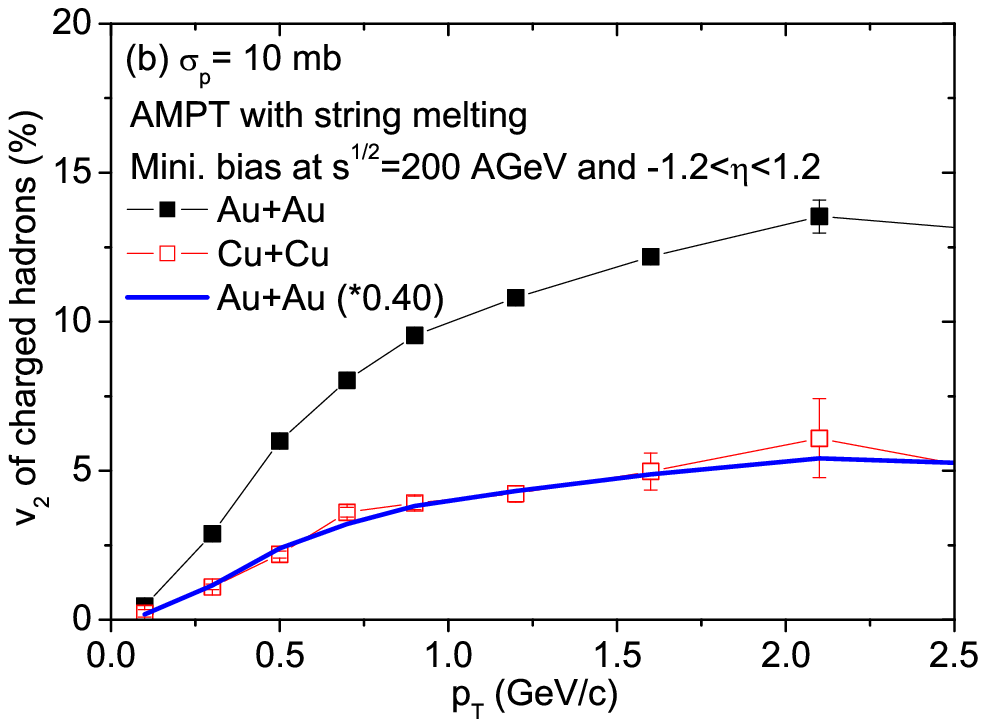}
\end{minipage}
\vspace{-0.5cm} \caption{Transverse momentum dependence of the
$v_{2}$ of mid-rapidity charged hadrons in minimum bias events 
of Au+Au (solid squares) and Cu+Cu (open squares) collisions at
$\sqrt{s}=200$ AGeV.}\label{cucu}
\end{figure}

In Fig. \ref{cucu}, results from the AMPT model with string melting
for charged hadron elliptic flows in minimum bias Cu+Cu and Au+Au
collisions at $\sqrt{s}=200$ AGeV are shown for parton scattering
cross sections of $\sigma _{p}=3$ mb (left panel) and 10 mb 
(right panel) \cite{chen3}.  It is seen that the elliptic flow
in the lighter Cu+Cu collisions is about a factor of 3 smaller than
that in the heavier Au+Au collisions at same energy as shown by
solid lines. This is consistent with the linear scaling of
the system size as well as the combined effect of the initial
energy density and spatial eccentricity.

\section{Anisotropic flows in collisions of asymmetric systems}

\vspace{-0.4cm}

\begin{figure}[htb]
\begin{minipage}{18pc}
\includegraphics[scale=0.65]{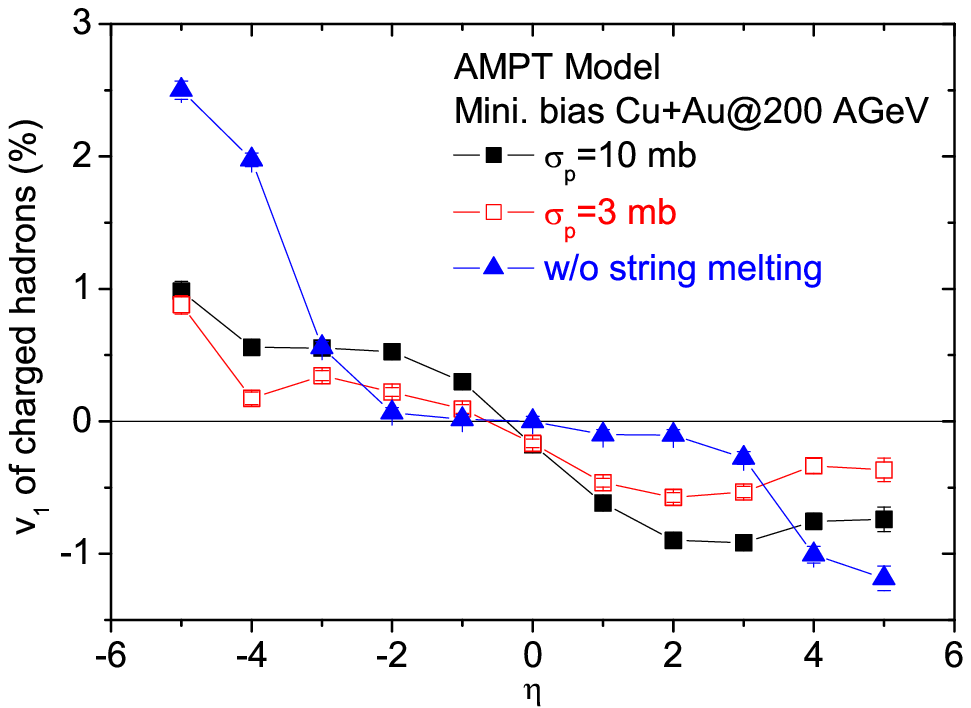}
\end{minipage}
\begin{minipage}{18pc}
\includegraphics[scale=0.65]{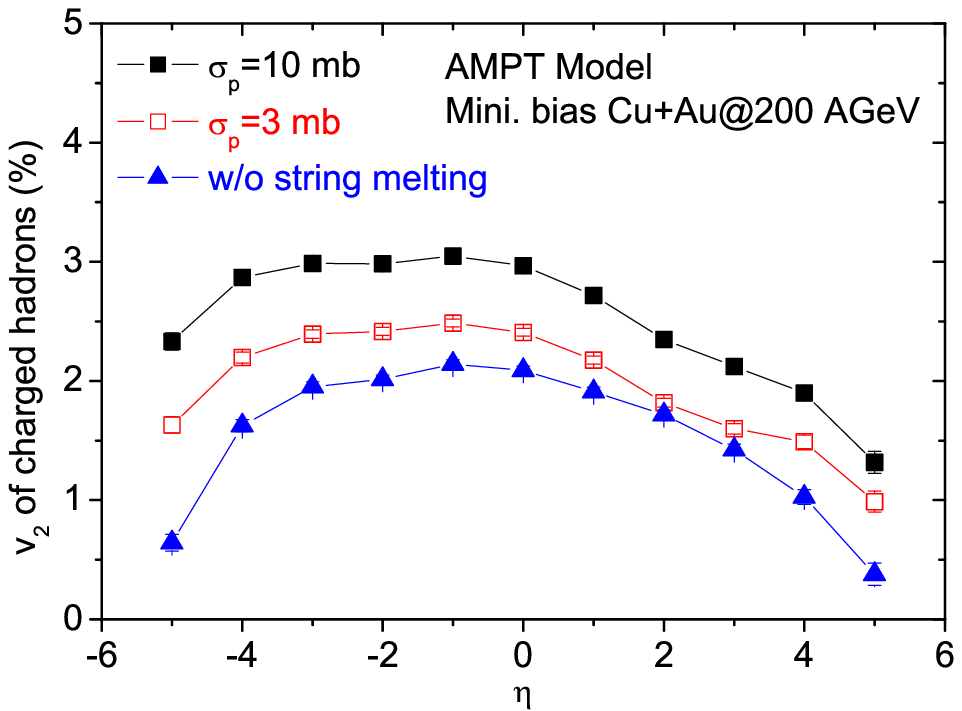}
\end{minipage}
\vspace{-0.5cm}
\caption{Pseudorapidity dependence of $v_1$ (left panel) and
$v_{2}$ (right panel) for charged hadrons in minimum bias events
of Cu+Au collisions at $\sqrt{s}=200$ AGeV.}
\label{cuau}
\end{figure}

In Fig. \ref{cuau}, the pseudorapidity dependence of $v_1$ (left 
panel) and $v_{2}$ (right panel) for charged hadrons in
minimum bias events of Cu+Au collisions at $\sqrt{s}=200$ AGeV are
shown for the string melting scenario with parton scattering cross
sections $\sigma _{p}=3$ (open squares) and $10$ mb (solid squares)
as well as the default AMPT model without string melting
(solid triangles) \cite{chen4}. Comparing with results
in symmetric Au+Au collisions, we find that charged hadrons produced around
mid-rapidity in asymmetric Cu+Au collisions display a stronger $v_1$
and their $v_2$ is also more sensitive to the parton cross
section used in the parton cascade. Furthermore, both $v_{1}$ and $v_{2}$
are appreciable and show an asymmetry in the forward and backward
rapidities.

\section{Summary}

Using the \textrm{AMPT} model, we have studied the rapidity and
colliding system size dependence of anisotropic flows in
heavy ion collisions at \textrm{RHIC}. We find that results on
the rapidity dependence of anisotropic flows suggest that a partonic
matter is formed during early stage of relativistic heavy ion
collisions only around mid-rapidity and that strings remain
dominant at large rapidities. Furthermore, to reproduce the
experimental data requires a parton cross section that is larger
than that given by the perturbative QCD, indicating that 
nonperterbative effects are important in the produced partonic
matter at RHIC. Also, a linear scaling with the
colliding system size is observed for the elliptic flow of
charged hadrons in minimum bias collisions. For collisions of
asymmetric systems, there is a strong $v_1$ around mid-rapidity,
and both $v_1$ and $v_2$ are asymmetric in the forward and backward 
rapidities. Experimental verification of latter predictions will be
very useful in testing the AMPT model as well as in understanding 
the dynamics of the partonic matter produced in the collisions.


\begin{thebibliography}{99}

\bibitem{Zhang99}B. Zhang, M. Gyulassy, and C. M. Ko, Phys. Lett. B 455
(1999) 45.

\bibitem{Zhang:1997ej}B. Zhang, Comput. Phys. Commun. 109 (1998) 193.

\bibitem{Zhang:2000bd}B. Zhang, C. M. Ko, B. A. Li, and Z. W. Lin,
Phys. Rev. C 61 (2000) 067901.

\bibitem{Lin:2001cx}Z. W. Lin, S. Pal, C. M. Ko, B. A. Li, and B.
Zhang, Phys. Rev. C 64 (2001) 011902 ; Nucl. Phys. A 698 (2002) 375;
nucl-th/0411110.

\bibitem{Lin:2001zk}Z. W. Lin and C. M. Ko, Phys. Rev. C 65 (2002) 034904.

\bibitem{chen1}L. W. Chen, C. M. Ko, and Z. W. Lin, Phys. Rev. C
69 (2004) 031901(R).

\bibitem{zhang}B. Zhang, L. W. Chen, and C. M. Ko, Phys. Rev. C 72 (2005)
024906.

\bibitem{chen2}L. W. Chen, V. Greco, C. M. Ko, and P. Kolb, Phys. Lett.
B 605 (2005) 95.

\bibitem{chen3}L. W. Chen and C. M. Ko, nucl-th/0505044.

\bibitem{chen4}L. W. Chen and C. M. Ko, nucl-th/0507067.

\bibitem{Wang:1991ht}X. N. Wang and M. Gyulassy, Phys. Rev. D 44 (1991)
3501.

\bibitem{Sjostrand:1994yb}T. Sjostrand, Comput. Phys. Commun. 82 (1994) 74.

\bibitem{Li:1995pr}B. A. Li and C. M. Ko, Phys. Rev. C 52 (1995)2037;
B. A. Li, A. T. Sustich, B. Zhang, and C.M. Ko, Int. Jour. Phys. E 10
(2001) 267.

\bibitem{STAR03}J. Adams {\it et al.} [STAR Collaboration], Phys. Rev.
Lett. 92 (2004) 062301.

\bibitem{manly03}S. Manly for the PHOBOS Collaboration, Nucl. Phys. A
715 (2003) 611c.

\bibitem{oldenburg04}M. D. Oldenburg for the STAR Collaboration,
2004, nucl-ex/0403007.

\end{thebibliography}
\end{document}